\documentclass[%
 reprint,
superscriptaddress,
 amsmath,amssymb,
 aps,
prl,
]{revtex4-2}

\usepackage{graphicx}
\usepackage{dcolumn}
\usepackage{bm}
\usepackage{hyperref}
\usepackage[mathlines]{lineno}

\usepackage{xcolor}

\usepackage[version=4]{mhchem}
\usepackage{physics}
\usepackage{siunitx}
\usepackage{orcidlink}

\newcommand{\bmb}{\ensuremath{B_{\mathrm{MB}}}}

\DeclareSIUnit\angstrom{\text{Å}}

\begin{document}

\title{Magnetic Field Reorganization of Electronic States in Moiré Bilayer Graphene}

\author{Milan Sharma \surname{Mandigo-Stoba}\,\orcidlink{0000-0003-1118-3132}}
\email{milanmsr@physics.ucla.edu}

\author{William Wang}%
\altaffiliation[Present address:]{
Department of Physics, University of California, Santa Barbara, CA 93106, USA.
}
\author{Jackson Kuklin}%
\author{Jack Lichterman}%
\altaffiliation[Present address:]{
Applied Materials, Santa Clara, CA 95054, USA.
}
\affiliation{%
 Department of Physics and Astronomy, University of California, Los Angeles, California 90095, USA
}%

\author{Kenji Watanabe\,\orcidlink{0000-0003-3701-8119}}%
\author{Takashi Taniguchi\,\orcidlink{0000-0002-1467-3105}}%
\affiliation{%
National Institute for Materials Science, Namiki 1-1, Tsukuba, Ibaraki 305-0044, Japan
}%

\author{Qianhui Shi\,\orcidlink{0000-0002-7606-3329}}%
 \email{shiqianhui@ucla.edu}
\affiliation{%
 Department of Physics and Astronomy, University of California, Los Angeles, California 90095, USA
}%

\date{\today}

\begin{abstract}
Magnetic fields are widely used to diagnose quantum phases in two-dimensional systems through quantum oscillations or by tuning spin and valley polarizations, but magnetic fields can also reshape the underlying electronic structure.
Here, using a bilayer graphene/hBN moiré system, we reveal a rich magnetic field induced evolution of the semiclassical orbit network, encompassing Lifshitz transitions, magnetic breakdown, and scattering between coexisting electron and hole pockets.
At magnetic fields of \SIrange{1}{2}{T}, quantum oscillation frequencies and the Hall density change markedly over a broad carrier density range, signaling magnetic breakdown and magnetic Lifshitz transitions. 
This evolution is valley contrasting: Berry curvature hot spots near the breakdown junctions enhance magnetic breakdown in the K valley while suppressing it in the K$^\prime$ valley, whereas valley-antisymmetric orbital magnetic moments split the corresponding Lifshitz transitions. 
The resulting valley-selective trajectories manifest at higher fields as valley-symmetry-breaking Hofstadter gaps. 
At elevated temperatures and low magnetic fields, scattering between coexisting electron and hole pockets produces nearly density-independent resistance oscillations whose frequency tracks the sum of their Fermi surface areas, persisting after conventional Onsager oscillations are thermally washed out.
Our results provide a unified picture of how modest magnetic fields reorganize moiré electronic states as the system evolves from semiclassical transport toward the Hofstadter regime.

\end{abstract}

\maketitle

Moiré superlattices, made from interfering lattices at 2D material interfaces, offer a versatile route to engineering band structures and open opportunities to explore new phases of matter.
The enlarged real space moiré periodicity folds the electronic bands into a reduced moiré Brillouin zone (BZ), leading to band flattening, gap opening, and reconstructed Fermi surfaces (FSs).
The interplay of magnetic fields and moiré superlattices is a rich subject, the most well-known example being the Hofstadter butterfly,
prominent when a rational fraction of a flux quantum pierces each unit cell~\cite{Hofstadter1976,Dean2013,Hunt2013,Ponomarenko2013}.
However, even before the Hofstadter regime, a magnetic field can significantly modify electronic states.

The magnetic field changes semiclassical trajectories through two mechanisms: orbital magnetic moments (OMMs), which shift underlying band energies, and magnetic breakdown (MB), which changes the connectivity of semiclassical orbits. 
A Bloch wavepacket's self-rotation generates OMM $\vb{m}(k)$, and a magnetic field modifies the band structure like $\varepsilon(k) \rightarrow \varepsilon(k) - \vb{m}(k)\vdot\vb{B}$~\cite{Xiao2010}.
A particularly prominent effect of OMMs on semiclassical trajectories occurs near a Van Hove singularity (VHS), across which the FS topology changes.
In this case, the magnetic field-induced band shift may drive the system through the fermiology change, producing a magnetic Lifshitz transition. 
On the other hand, even when the static FS topology is unchanged, the magnetic field may induce tunneling between adjacent Fermi pockets (FPs). 
While carriers at zero magnetic field trace their FPs in momentum space, at finite magnetic field the electronic wavefunctions acquire a finite extent in momentum space of order $1/l_B \propto \sqrt{B}$, with $l_B$ the magnetic length.
If nearby FPs are separated by $1/l_B$, tunneling between them is appreciable, creating new orbits and new quantum oscillation (QO) frequencies~\cite{Cohen1961,Blount1962,Chambers1966,Alexandradinata2017_1}.
The tunneling probability is $P = \exp(-\bmb/B)$, with $\bmb$ the breakdown field.
The post-MB orbits coexist with the original orbits; their interference ultimately evolves into magnetic minibands and the Hofstadter spectrum~\cite{Pippard1964,Machida1995,paul2022,deVries2024_1}. 

Moiré electronic structures are particularly susceptible to modification by modest magnetic fields:  their small BZs naturally favor MB, while their narrow energy scales make the effects of OMMs particularly consequential.
At the same time, QOs are widely used to characterize FSs and flavor degeneracy, often serving as key diagnostics of quantum phases.
It is therefore important to understand how magnetic fields reorganize electronic states. 
Despite a few recent studies~\cite{Moon2024_1,bocarsly2024_1,han2025}, the transport signatures of MB and magnetic Lifshitz transitions in moiré systems remain largely unexplored.
Moreover, although MB was extensively studied in conventional metals in the 1960s~\cite{pippard1962,Priestley1963,Watts1964}, its interplay with Berry curvature has not been experimentally investigated.
Conventionally, MB is controlled by the FS geometry, including separation between FPs and local band velocities. 
However, the quantum geometry, which controls the overlap of Bloch wavefunctions, could also be important in modifying tunneling between FPs and therefore enhancing or suppressing MB.

\begin{figure*}
\includegraphics{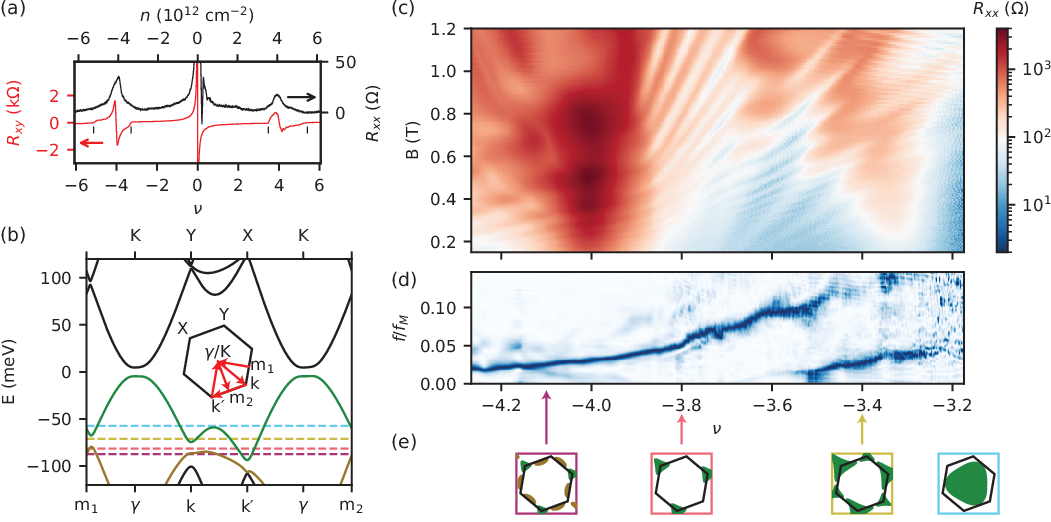}
\caption{\label{fig:1}
(a) Left (right), in red (black), is antisymmetrized Hall (longitudinal) resistance at $B = \SI{\pm0.3}{T}$ (\SI{0}{T}) and $D = \SI{0}{V.nm^{-1}}$. VHSs are marked by short black lines.
(b) Calculated low energy band structure of moiré BLG/hBN at $\theta=\ang{0.864}$ \cite{SuppMat}. Inset: path through moiré BZ, with fermi energies in (e) indicated by dashed lines.
(c) Low field longitudinal resistance map around $\nu = -4$ at $D = \SI{-0.20}{V.nm^{-1}}$.
(d) Fourier transform of $\dv*{R_{xx}(1/B)}{B}$ below \SI{0.6}{T}, $f/f_M = f/(\phi_0 n_M)$.
(e) FSs at $\nu = \SIlist{-4.1;-3.8;-3.4;-2.5}{}$. 
FPs from the first (second) moiré valence band are shown in green (dark yellow).}
\end{figure*}

Here, we study the effects of magnetic fields on moiré materials in a high-quality bilayer graphene/hexagonal boron nitride (BLG/hBN) moiré superlattice.
Our data span three regimes as the magnetic field increases, showing the evolution from the low field fermiology of the moiré minibands, through MB and magnetic Lifshitz transitions, to the Hofstadter spectrum.
At low magnetic fields below \SI{1}{T}, we resolve QOs to map the multi-pocket and multi-band fermiology of moiré BLG/hBN in the first and second moiré valence bands.
At modest magnetic fields of \SIrange{1}{2}{T}, MB and magnetic Lifshitz transitions change the QO frequencies and Hall density $n_H = B/eR_{xy}$ in a wide density range.
Furthermore, we find that MB junctions naturally appear at Berry curvature hot spots; the Berry curvature and OMM of the moiré minibands further modify the local band geometry and tunneling conditions.
The valley-antisymmetric nature of the Berry curvature and OMM imparts a valley-contrasting character onto MB and magnetic Lifshitz transitions.
Elevated temperatures further reveal nearly density-independent QOs at magnetic fields below \SI{1}{T} due to the scattering between coexisting electron and hole pockets.
Our results illustrate how magnetic fields can reshape bands and modify transport properties in a moiré system, highlighting the interplay with quantum geometry, and demonstrate the need for careful interpretation of QOs.

In Fig.~\ref{fig:1}, we present an overview of the electronic properties of BLG/BN with a twist angle of \ang{0.86} at zero and low magnetic fields.
\autoref{fig:1}~(a) shows the zero-field longitudinal resistance and low-field Hall resistance. Weakly resistive hole and electron satellite peaks accompanied by a diverging Hall resistance suggest the separation of a moiré band at $\nu = \pm 4$.
Around $\nu = \SIlist{-5.11;-3.26;3.47;5.38}{}$, $R_{xy}$ changes sign, indicating several VHSs.
The low-energy band structure is shown in Fig.~\ref{fig:1}~(b), with $E_F$ indicated for the FSs shown in Fig.~\ref{fig:1}~(e).
A detailed map of QOs around $\nu = -4$ at low magnetic fields and its Fourier transform, shown in Figs.~\ref{fig:1}~(c) and (d), reveal the details of the fermiology:
At the 1st moiré valence band minimum, two electron pockets open around $\nu = -4.7$ (the X pocket) and $\nu = -3.7$ (the Y pocket), using notation from Ref.~\cite{moon2014}.
The X electron pocket overlaps with three hole pockets at the 2nd moiré valence band maximum, changing the slope of the X pocket Landau fan.
Coexisting electron and hole pockets, rather than a global gap opening at $\nu = -4$, are consistent with previous theoretical and experimental works~\cite{bocarsly2024_1}.

\begin{figure*}[htb]
\includegraphics{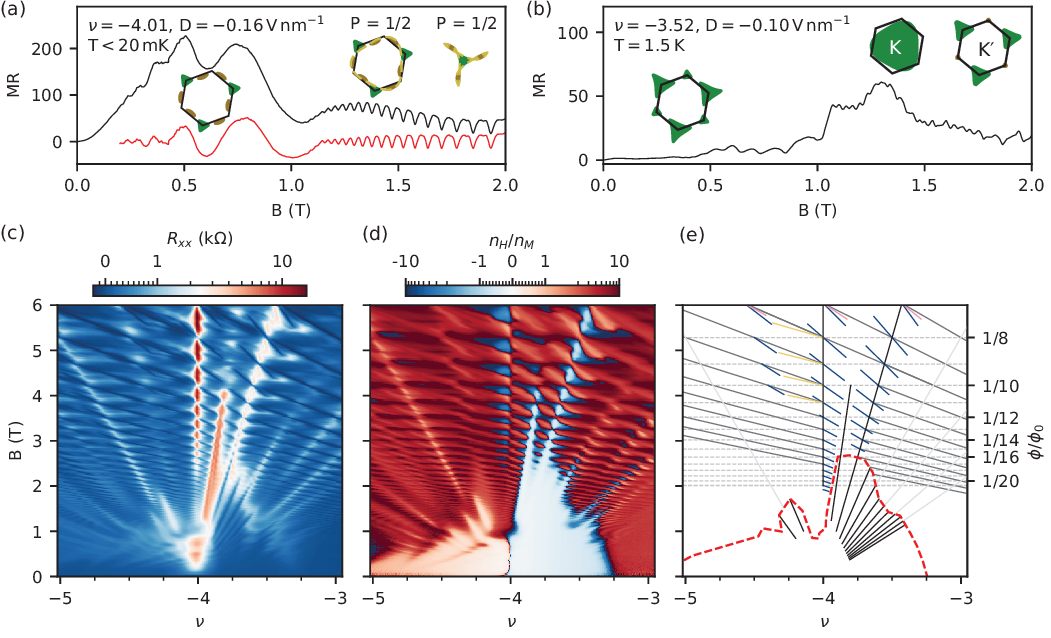}
\caption{\label{fig:2} (a) Symmetrized MR at $n=\SI{-4.097e12}{cm^{-2}}$ in black and background subtracted MR in red.
Insets: schematic of low field orbits and high field MB orbits, with electron (hole) pockets in green (dark yellow), and MB orbits in yellow.
(b) Symmetrized MR at $n = \SI{-3.60e12}{cm^{-2}}$.
Insets: schematic of low field FS and high field valley-resolved FSs.
(c) Symmetrized longitudinal resistance and (d) antisymmetrized Hall density map with $D=\SI{-0.10}{V.nm^{-1}}$, measured at \SI{1.5}{K}.
(e) Wannier diagram with moiré LL gaps in black, their extension into the MB/magnetic Lifshitz transition regime in light gray, bilayer graphene LL gaps in dark gray and symmetry-broken Hofstadter states emerging from $\nu=\SIlist{-2;2;-1}{}$ in blue, yellow, and red, respectively. Brown-Zak oscillations with $\phi/\phi_0 = 1/q$ are drawn as dashed horizontal lines~\cite{Brown1964,Zak1964}.}
\end{figure*}

A moderate magnetic field changes QO frequencies and $n_H$ through a magnetic Lifshitz transition and MB. 
\autoref{fig:2}~(a) shows the magnetoresistance (MR), $R_{xx}(B)/R_{xx}(0) - 1$, versus magnetic field for $\nu = -4.01$.
The system exhibits extremely large magnetoresistance (XMR) in excess of \SI{20000}{\percent}, emblematic of compensated electron and hole pockets~\cite{shilov2024,wang2023openorbitinducedlowfield}.
In addition, low frequency QOs indicating a FP size of \SI{0.032}{\text{$n_M$}} are observed, better seen in the red trace which shows the MR after the XMR background is subtracted.
Above \SI{1}{T}, XMR is suppressed and higher frequency QOs dominate, indicating a FP size of $\SI{1.02}{\text{$n_M$}}$.
At the same time, the Hall coefficient changes sign \cite{SuppMat}.
The change of QO frequency and Hall sign, concomitant with the suppression of XMR, indicates MB in the moiré bands.
Below 1 T, the carriers populate both the electron and the hole pockets simultaneously.
Above 1 T, the carriers tunnel from one FP to another, tracing out both the grand hole orbit of the 
pristine BLG valence band
and a mixed electron-hole orbit corresponding to a density of $n-4n_M$, as illustrated in the insets of Fig.~\ref{fig:2}~(a).

\autoref{fig:2}~(b) shows MR versus magnetic field at $\nu = -3.52$.
Again, below \SI{1}{T} low frequency QOs are observed, indicating a FP size of \SI{0.124}{\text{$n_M$}}, while above \SI{1}{T}, high frequency QOs dominate, indicating a FP size of \SI{0.88}{\text{$n_M$}}.
Near the VHS, intraband MB can also occur between the X and Y electron pockets, but the tunneling probability $P \leq 1/2$, with equality at the VHS~\cite{Alexandradinata2017_1,Kemble1935}.
Instead, the dominant effect is a valley-selective magnetic Lifshitz transition, as OMMs with opposite sign in K and K$^\prime$ split the VHS.
At moderate fields, K and K$^\prime$ host carriers of different type, causing $n_H$ to change and high frequency QOs to dominate.
Indeed, transport signatures of MB and magnetic Lifshitz transitions are observed in a wide range of densities between the VHS and $\nu = -5$, as shown in the maps of longitudinal resistance and $n_H$ versus moiré filling and magnetic field in Fig.~\ref{fig:2}~(c) and (d).
MB and magnetic Lifshitz transitions manifest in transport through three signatures:
(1) the QO frequency changes between low and high magnetic fields, 
(2) $n_H$ jumps across the same boundary, indicating a change in carrier number or type
and (3) an $R_{xx}$ maxima traces the boundary.
The schematic boundary is shown as the red dashed curve in Fig.~\ref{fig:2}~(e), with solid lines indicating resistance or conductance minima corresponding to Landau levels (LLs).

\begin{figure}
\includegraphics{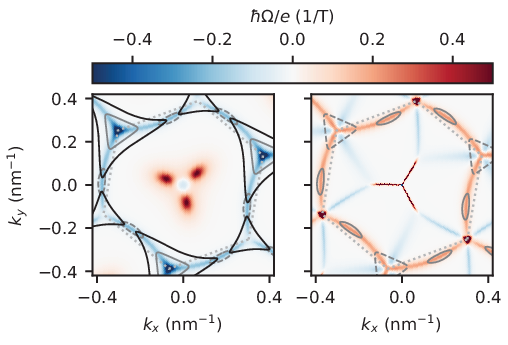}
\caption{\label{fig:BC} Left (right) shows Berry curvature in the first (second) moiré valence band in the K valley~\cite{SuppMat}. FSs at $\nu = -3.3$ ($-3.95$) overlaid in black (gray). Solid lines indicate FPs in that band, dashed lines indicate FPs in the other band.}
\end{figure}
Both the magnetic Lifshitz transition and MB are strongly affected by Berry curvature.
\autoref{fig:BC} shows the Berry curvature of the first two moiré valence bands in the K valley, with superimposed FSs at $\nu=-3.3$ and $-3.95$. 
The Berry curvature hot spots forms a ring, dubbed the ``Berry ring of fire'' in other graphene systems~\cite{Patri2025,Sheekey2026}, though the origin here slightly differs.
In BLG/hBN the moiré potential is an essential ingredient, reconstructing the bands and concentrating Berry curvature, whereas in rhombohedral graphene a large displacement field
is necessary~\cite{Patri2025}.
The MB junctions, where neighboring FPs are closest, naturally occur on the hot spot ring, as large Berry curvature is generated near small band gaps.
The OMM shifts band energies in a magnetic field and can drive the system through a Lifshitz transition.
Even when the shift is insufficient to change the static FS topology, it can reduce the separation between neighboring FPs and enhance MB.
In addition, Berry curvature modifies the semiclassical equations of motion, in particular, the rate at which wavepackets traverse momentum space, $\dv*{k}{t}$, thereby changing the tunneling probability~\cite{Xiao2010}.
Because the Berry curvature and OMM are opposite in the K and K$^\prime$ valleys, the band shifts and MB dynamics are valley-contrasting, leading to different MB and magnetic Lifshitz transitions in the two valleys.

The valley-contrasting MB and magnetic Lifshitz transitions are reflected in the low $\bmb$ in a large density range: the boundary outlined in Fig.~\ref{fig:2}~(e) indicates the smaller of $\bmb$ in K and $\bmb$ in K$^\prime$.
Other experimental signatures also attest to the valley-contrasting character.
First, while the low field QOs from the X electron pocket (black lines in Fig.~\ref{fig:2}~(e)) exhibit a degeneracy of two, showing resistance minima when the pocket in either the K or K$^\prime$ valley is at integer filling $\nu_{eLL} = (4n_M +n)h/eB = 2j$ with integer $j$, after the magnetic Lifshitz transition, the conductance minima following an electron-type Landau fan (light gray lines in Fig.~\ref{fig:2}~(e)) only survive at $\nu_{eLL} = 4j$.
This apparent periodicity doubling is consistent with a scenario where carriers populate both valleys, and one valley produces an electron-type Landau fan, while the other valley undergoes a magnetic Lifshitz transition, producing a hole-type Landau fan.
Second, as seen from Figs.~\ref{fig:2}~(c)-(e), at $\nu$ close to $-4$, the post-MB oscillations---with resistance minima indicated by dark blue segments in Fig.~\ref{fig:2}~(e)---give a FP size of $n - 2n_M$ and degeneracy of 2, suggesting a fully valley-polarized FS.
This is consistent with MB that only happens in one valley, while the other valley is gapped.
The valley-contrasting post-MB orbits coexist with the pre-MB hole orbits over a broad field range. 
As can be seen from Figs.~\ref{fig:2}~(c)-(e), the QOs originating from both sets of orbits overlap, and their interplay evolves into the Hofstadter spectrum at high fields.

\begin{figure}
\includegraphics{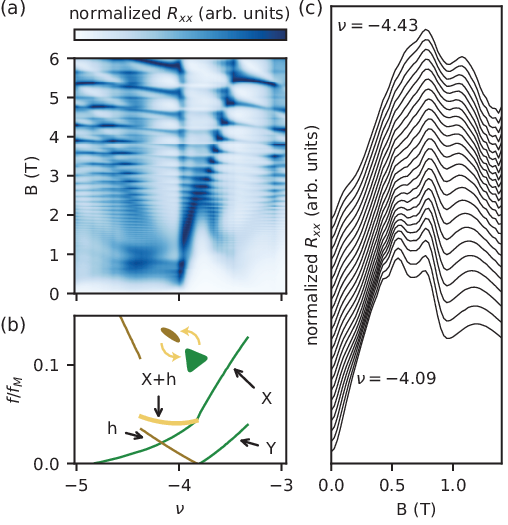}
\caption{
\label{fig:Temp} (a)
Symmetrized longitudinal resistance map around $\nu=-4$ with $D=\SI{-0.10}{V.nm^{-1}}$, measured at \SI{18}{K}. At each density, $R_{xx}$ is divided by the maximum value of $R_{xx}$, to enhance visibility of low field QOs.
(b)
Calculated QO frequencies for the various FPs. Inset shows the X electron and lobe hole pockets that make the $\text{X}+\text{h}$ frequency.
(c)
Symmetrized longitudinal resistance line cuts for $\nu \in [-4.08, -4.43]$, averaged over $\var\nu = 0.01$ and offset for clarity. At each density, $R_{xx}$ is divided by the maximum value of $R_{xx}$ for $B < \SI{1}{T}$, to enhance visibility of low field QOs.}
\end{figure}

Elevated temperatures reveal new QOs.
\autoref{fig:Temp}~(a) shows the longitudinal resistance at \SI{18}{K}, where low-field LLs from the X and Y electron pockets are notably absent.
At high fields, the $\nu_{eLL} = +4$ LL persists, as well as the hole-like LLs from the first valence band.
Density-independent oscillations, with frequency $\phi_0 n_M$, are also visible at high fields.
Such oscillations may arise either from the recurring formation of extended magnetic Bloch minibands at rational values of magnetic flux quanta per moiré unit cell, known as Brown–Zak oscillations \cite{Brown1964,Zak1964,krishnakumar2017}, or from Aharonov–Bohm interference between alternative paths in a network of electron- and hole-like semiclassical trajectories near the Lifshitz transition, analogous to the kagome oscillations reported in Ref.~\citep{deVries2024_1}.
The high field density-independent oscillations observed here likely lie in the crossover between these two limits~\cite{deVries2024_1}.

At low fields, we observe a distinct set of nearly density-independent QOs between $\nu=-4.5$ and $\nu=-4$ [See also Fig.~4~(c)], corresponding to a FP size of $\approx \SI{0.05}{\text{$n_M$}}$, consistent with the calculated sum of the X electron pocket and the threefold degenerate ``lobe'' hole pocket from the second valence band, as shown in Fig.~4~(b).
Above \SI{1.2}{T}, these QOs are suppressed as MB becomes dominant, seen in the coinciding jump in $n_H$~\cite{SuppMat}.
We attribute these oscillations to quasiparticle lifetime oscillations (QPLO)~\cite{Huber2023,Leeb2025}---a more general form of magneto-intersubband/intervalley oscillations (MISO)~\cite{leadley1992,raikh1994,Phinney2021}---which, in this case, arises from the scattering between one electron pocket and one hole pocket.
Because the interpocket scattering rate depends on the product of the oscillatory densities of states of the two pockets with frequencies $f_1$ and $f_2$, the resistance can contain both sum and difference components $f_1+f_2$ and $|f_1-f_2|$. 
In the more-studied case where the scattering involves two pockets of the same carrier type, the thermally robust component is the difference component $|f_1-f_2|$;
in the present case, by contrast, scattering between the electron and hole pockets generates the nearly density-independent frequency $f_X + f_h$, with effective mass that controls thermal damping $m_X^* + m_h^* = \abs{m_X^*} - \abs{m_h^*} \ll \abs{m_X^*}, \abs{m_h^*}$.
The reduced effective mass is consistent with our observation of these QOs at elevated temperatures and low fields, where the X electron LLs (with cyclotron mass $m_X^*$) are thermally washed out
\footnote{
Increased temperature smears the FS, and QOs with cyclotron mass $m^* = \hbar^{-2} \dv*{A_k}{\mu}$ (with $A_k$ the extremal FS cross-section area)
are damped according to the well-known Lifshitz-Kosevich thermal damping factor $R_T = X/\sinh{X}$, with $X\propto m^* T / B$ \cite{Lifshitz}.
Density-independent QO frequencies, with near-zero $m^*$, can survive to much higher temperatures~\cite{deVries2024_1,Leeb2025}.
}.
On the other hand, MB would produce a density-dependent difference frequency $f_X - f_h$ and cannot account for the observed oscillations.
More broadly, interpocket scattering provides a third route by which multiband fermiology gives rise to unconventional quantum oscillations, in addition to magnetic Lifshitz transitions and MB.

To conclude, we have systematically studied how a modest magnetic field reorganizes the moiré fermiology in a BLG/hBN heterostructure and connects it to the Hofstadter regime.
The correlated changes in quantum oscillations and Hall response show that the magnetic field can modify the electronic states through OMM-induced changes in the minibands, magnetic breakdown and scattering between different pockets, rather than simply probe the zero-field Fermi surface. 
Similar effects should be broadly relevant in other flat-band systems \cite{lu2014,kalantre2026,zhao2026}, where small momentum-space separations and strong orbital coupling to magnetic field make the electronic structure particularly susceptible to modest fields.

\begin{acknowledgments}
The authors thank Pilkyung Moon for helpful discussions.
A portion of this work was performed at the National High Magnetic Field Laboratory, which is supported by National Science Foundation Cooperative Agreement No. DMR-2128556 and the State of Florida.    
\end{acknowledgments}


\begin{thebibliography}{39}%
\makeatletter
\providecommand \@ifxundefined [1]{%
 \@ifx{#1\undefined}
}%
\providecommand \@ifnum [1]{%
 \ifnum #1\expandafter \@firstoftwo
 \else \expandafter \@secondoftwo
 \fi
}%
\providecommand \@ifx [1]{%
 \ifx #1\expandafter \@firstoftwo
 \else \expandafter \@secondoftwo
 \fi
}%
\providecommand \natexlab [1]{#1}%
\providecommand \enquote  [1]{``#1''}%
\providecommand \bibnamefont  [1]{#1}%
\providecommand \bibfnamefont [1]{#1}%
\providecommand \citenamefont [1]{#1}%
\providecommand \href@noop [0]{\@secondoftwo}%
\providecommand \href [0]{\begingroup \@sanitize@url \@href}%
\providecommand \@href[1]{\@@startlink{#1}\@@href}%
\providecommand \@@href[1]{\endgroup#1\@@endlink}%
\providecommand \@sanitize@url [0]{\catcode `\\12\catcode `\$12\catcode `\&12\catcode `\#12\catcode `\^12\catcode `\_12\catcode `\%12\relax}%
\providecommand \@@startlink[1]{}%
\providecommand \@@endlink[0]{}%
\providecommand \url  [0]{\begingroup\@sanitize@url \@url }%
\providecommand \@url [1]{\endgroup\@href {#1}{\urlprefix }}%
\providecommand \urlprefix  [0]{URL }%
\providecommand \Eprint [0]{\href }%
\providecommand \doibase [0]{https://doi.org/}%
\providecommand \selectlanguage [0]{\@gobble}%
\providecommand \bibinfo  [0]{\@secondoftwo}%
\providecommand \bibfield  [0]{\@secondoftwo}%
\providecommand \translation [1]{[#1]}%
\providecommand \BibitemOpen [0]{}%
\providecommand \bibitemStop [0]{}%
\providecommand \bibitemNoStop [0]{.\EOS\space}%
\providecommand \EOS [0]{\spacefactor3000\relax}%
\providecommand \BibitemShut  [1]{\csname bibitem#1\endcsname}%
\let\auto@bib@innerbib\@empty
\bibitem [{\citenamefont {Hofstadter}(1976)}]{Hofstadter1976}%
  \BibitemOpen
  \bibfield  {author} {\bibinfo {author} {\bibfnamefont {D.~R.}\ \bibnamefont {Hofstadter}},\ }\href {https://doi.org/10.1103/PhysRevB.14.2239} {\bibfield  {journal} {\bibinfo  {journal} {Phys. Rev. B}\ }\textbf {\bibinfo {volume} {14}},\ \bibinfo {pages} {2239} (\bibinfo {year} {1976})}\BibitemShut {NoStop}%
\bibitem [{\citenamefont {{Dean}}\ \emph {et~al.}(2013)\citenamefont {{Dean}}, \citenamefont {{Wang}}, \citenamefont {{Maher}}, \citenamefont {{Forsythe}}, \citenamefont {{Ghahari}}, \citenamefont {{Gao}}, \citenamefont {{Katoch}}, \citenamefont {{Ishigami}}, \citenamefont {{Moon}}, \citenamefont {{Koshino}},\ and\ \citenamefont {et~al.}}]{Dean2013}%
  \BibitemOpen
  \bibfield  {author} {\bibinfo {author} {\bibfnamefont {C.~R.}\ \bibnamefont {{Dean}}}, \bibinfo {author} {\bibfnamefont {L.}~\bibnamefont {{Wang}}}, \bibinfo {author} {\bibfnamefont {P.}~\bibnamefont {{Maher}}}, \bibinfo {author} {\bibfnamefont {C.}~\bibnamefont {{Forsythe}}}, \bibinfo {author} {\bibfnamefont {F.}~\bibnamefont {{Ghahari}}}, \bibinfo {author} {\bibfnamefont {Y.}~\bibnamefont {{Gao}}}, \bibinfo {author} {\bibfnamefont {J.}~\bibnamefont {{Katoch}}}, \bibinfo {author} {\bibfnamefont {M.}~\bibnamefont {{Ishigami}}}, \bibinfo {author} {\bibfnamefont {P.}~\bibnamefont {{Moon}}}, \bibinfo {author} {\bibfnamefont {M.}~\bibnamefont {{Koshino}}},\ and\ \bibinfo {author} {\bibnamefont {et~al.}},\ }\href {https://doi.org/10.1038/nature12186} {\bibfield  {journal} {\bibinfo  {journal} {\nat}\ }\textbf {\bibinfo {volume} {497}},\ \bibinfo {pages} {598} (\bibinfo {year} {2013})}\BibitemShut {NoStop}%
\bibitem [{\citenamefont {Hunt}\ \emph {et~al.}(2013)\citenamefont {Hunt}, \citenamefont {Sanchez-Yamagishi}, \citenamefont {Young}, \citenamefont {Yankowitz}, \citenamefont {LeRoy}, \citenamefont {Watanabe}, \citenamefont {Taniguchi}, \citenamefont {Moon}, \citenamefont {Koshino}, \citenamefont {Jarillo-Herrero},\ and\ \citenamefont {Ashoori}}]{Hunt2013}%
  \BibitemOpen
  \bibfield  {author} {\bibinfo {author} {\bibfnamefont {B.}~\bibnamefont {Hunt}}, \bibinfo {author} {\bibfnamefont {J.~D.}\ \bibnamefont {Sanchez-Yamagishi}}, \bibinfo {author} {\bibfnamefont {A.~F.}\ \bibnamefont {Young}}, \bibinfo {author} {\bibfnamefont {M.}~\bibnamefont {Yankowitz}}, \bibinfo {author} {\bibfnamefont {B.~J.}\ \bibnamefont {LeRoy}}, \bibinfo {author} {\bibfnamefont {K.}~\bibnamefont {Watanabe}}, \bibinfo {author} {\bibfnamefont {T.}~\bibnamefont {Taniguchi}}, \bibinfo {author} {\bibfnamefont {P.}~\bibnamefont {Moon}}, \bibinfo {author} {\bibfnamefont {M.}~\bibnamefont {Koshino}}, \bibinfo {author} {\bibfnamefont {P.}~\bibnamefont {Jarillo-Herrero}},\ and\ \bibinfo {author} {\bibfnamefont {R.~C.}\ \bibnamefont {Ashoori}},\ }\href {https://doi.org/10.1126/science.1237240} {\bibfield  {journal} {\bibinfo  {journal} {Science}\ }\textbf {\bibinfo {volume} {340}},\ \bibinfo {pages} {1427} (\bibinfo {year} {2013})}\BibitemShut {NoStop}%
\bibitem [{\citenamefont {{Ponomarenko}}\ \emph {et~al.}(2013)\citenamefont {{Ponomarenko}}, \citenamefont {{Gorbachev}}, \citenamefont {{Yu}}, \citenamefont {{Elias}}, \citenamefont {{Jalil}}, \citenamefont {{Patel}}, \citenamefont {{Mishchenko}}, \citenamefont {{Mayorov}}, \citenamefont {{Woods}}, \citenamefont {{Wallbank}},\ and\ \citenamefont {et~al.}}]{Ponomarenko2013}%
  \BibitemOpen
  \bibfield  {author} {\bibinfo {author} {\bibfnamefont {L.~A.}\ \bibnamefont {{Ponomarenko}}}, \bibinfo {author} {\bibfnamefont {R.~V.}\ \bibnamefont {{Gorbachev}}}, \bibinfo {author} {\bibfnamefont {G.~L.}\ \bibnamefont {{Yu}}}, \bibinfo {author} {\bibfnamefont {D.~C.}\ \bibnamefont {{Elias}}}, \bibinfo {author} {\bibfnamefont {R.}~\bibnamefont {{Jalil}}}, \bibinfo {author} {\bibfnamefont {A.~A.}\ \bibnamefont {{Patel}}}, \bibinfo {author} {\bibfnamefont {A.}~\bibnamefont {{Mishchenko}}}, \bibinfo {author} {\bibfnamefont {A.~S.}\ \bibnamefont {{Mayorov}}}, \bibinfo {author} {\bibfnamefont {C.~R.}\ \bibnamefont {{Woods}}}, \bibinfo {author} {\bibfnamefont {J.~R.}\ \bibnamefont {{Wallbank}}},\ and\ \bibinfo {author} {\bibnamefont {et~al.}},\ }\href {https://doi.org/10.1038/nature12187} {\bibfield  {journal} {\bibinfo  {journal} {\nat}\ }\textbf {\bibinfo {volume} {497}},\ \bibinfo {pages} {594} (\bibinfo {year} {2013})}\BibitemShut {NoStop}%
\bibitem [{\citenamefont {Xiao}\ \emph {et~al.}(2010)\citenamefont {Xiao}, \citenamefont {Chang},\ and\ \citenamefont {Niu}}]{Xiao2010}%
  \BibitemOpen
  \bibfield  {author} {\bibinfo {author} {\bibfnamefont {D.}~\bibnamefont {Xiao}}, \bibinfo {author} {\bibfnamefont {M.-C.}\ \bibnamefont {Chang}},\ and\ \bibinfo {author} {\bibfnamefont {Q.}~\bibnamefont {Niu}},\ }\href {https://doi.org/10.1103/RevModPhys.82.1959} {\bibfield  {journal} {\bibinfo  {journal} {Rev. Mod. Phys.}\ }\textbf {\bibinfo {volume} {82}},\ \bibinfo {pages} {1959} (\bibinfo {year} {2010})}\BibitemShut {NoStop}%
\bibitem [{\citenamefont {Cohen}\ and\ \citenamefont {Falicov}(1961)}]{Cohen1961}%
  \BibitemOpen
  \bibfield  {author} {\bibinfo {author} {\bibfnamefont {M.~H.}\ \bibnamefont {Cohen}}\ and\ \bibinfo {author} {\bibfnamefont {L.~M.}\ \bibnamefont {Falicov}},\ }\href {https://doi.org/10.1103/PhysRevLett.7.231} {\bibfield  {journal} {\bibinfo  {journal} {Phys. Rev. Lett.}\ }\textbf {\bibinfo {volume} {7}},\ \bibinfo {pages} {231} (\bibinfo {year} {1961})}\BibitemShut {NoStop}%
\bibitem [{\citenamefont {Blount}(1962)}]{Blount1962}%
  \BibitemOpen
  \bibfield  {author} {\bibinfo {author} {\bibfnamefont {E.~I.}\ \bibnamefont {Blount}},\ }\href {https://doi.org/10.1103/PhysRev.126.1636} {\bibfield  {journal} {\bibinfo  {journal} {Phys. Rev.}\ }\textbf {\bibinfo {volume} {126}},\ \bibinfo {pages} {1636} (\bibinfo {year} {1962})}\BibitemShut {NoStop}%
\bibitem [{\citenamefont {Chambers}(1966)}]{Chambers1966}%
  \BibitemOpen
  \bibfield  {author} {\bibinfo {author} {\bibfnamefont {R.~G.}\ \bibnamefont {Chambers}},\ }\href {https://doi.org/10.1088/0370-1328/88/3/318} {\bibfield  {journal} {\bibinfo  {journal} {Proc. Phys. Soc., London}\ }\textbf {\bibinfo {volume} {88}},\ \bibinfo {pages} {701} (\bibinfo {year} {1966})}\BibitemShut {NoStop}%
\bibitem [{\citenamefont {Alexandradinata}\ and\ \citenamefont {Glazman}(2017)}]{Alexandradinata2017_1}%
  \BibitemOpen
  \bibfield  {author} {\bibinfo {author} {\bibfnamefont {A.}~\bibnamefont {Alexandradinata}}\ and\ \bibinfo {author} {\bibfnamefont {L.}~\bibnamefont {Glazman}},\ }\href {https://doi.org/10.1103/PhysRevLett.119.256601} {\bibfield  {journal} {\bibinfo  {journal} {Phys. Rev. Lett.}\ }\textbf {\bibinfo {volume} {119}},\ \bibinfo {pages} {256601} (\bibinfo {year} {2017})}\BibitemShut {NoStop}%
\bibitem [{\citenamefont {Pippard}(1964)}]{Pippard1964}%
  \BibitemOpen
  \bibfield  {author} {\bibinfo {author} {\bibfnamefont {A.~B.}\ \bibnamefont {Pippard}},\ }\href {https://doi.org/10.1098/rsta.1964.0008} {\bibfield  {journal} {\bibinfo  {journal} {Philos. Trans. R. Soc., A}\ }\textbf {\bibinfo {volume} {256}},\ \bibinfo {pages} {317} (\bibinfo {year} {1964})}\BibitemShut {NoStop}%
\bibitem [{\citenamefont {Machida}\ \emph {et~al.}(1995)\citenamefont {Machida}, \citenamefont {Kishigi},\ and\ \citenamefont {Hori}}]{Machida1995}%
  \BibitemOpen
  \bibfield  {author} {\bibinfo {author} {\bibfnamefont {K.}~\bibnamefont {Machida}}, \bibinfo {author} {\bibfnamefont {K.}~\bibnamefont {Kishigi}},\ and\ \bibinfo {author} {\bibfnamefont {Y.}~\bibnamefont {Hori}},\ }\href {https://doi.org/10.1103/PhysRevB.51.8946} {\bibfield  {journal} {\bibinfo  {journal} {Phys. Rev. B}\ }\textbf {\bibinfo {volume} {51}},\ \bibinfo {pages} {8946} (\bibinfo {year} {1995})}\BibitemShut {NoStop}%
\bibitem [{\citenamefont {Paul}\ \emph {et~al.}(2022)\citenamefont {Paul}, \citenamefont {Crowley}, \citenamefont {Devakul},\ and\ \citenamefont {Fu}}]{paul2022}%
  \BibitemOpen
  \bibfield  {author} {\bibinfo {author} {\bibfnamefont {N.}~\bibnamefont {Paul}}, \bibinfo {author} {\bibfnamefont {P.~J.~D.}\ \bibnamefont {Crowley}}, \bibinfo {author} {\bibfnamefont {T.}~\bibnamefont {Devakul}},\ and\ \bibinfo {author} {\bibfnamefont {L.}~\bibnamefont {Fu}},\ }\href {https://doi.org/10.1103/PhysRevLett.129.116804} {\bibfield  {journal} {\bibinfo  {journal} {Phys. Rev. Lett.}\ }\textbf {\bibinfo {volume} {129}},\ \bibinfo {pages} {116804} (\bibinfo {year} {2022})}\BibitemShut {NoStop}%
\bibitem [{\citenamefont {de~Vries}\ \emph {et~al.}(2024)\citenamefont {de~Vries}, \citenamefont {Slizovskiy}, \citenamefont {Tomić}, \citenamefont {Krishna~Kumar}, \citenamefont {Garcia-Ruiz}, \citenamefont {Zheng}, \citenamefont {Portol{\'e}s}, \citenamefont {Ponomarenko}, \citenamefont {Geim}, \citenamefont {Watanabe}, \citenamefont {Taniguchi}, \citenamefont {Fal’ko}, \citenamefont {Ensslin}, \citenamefont {Ihn},\ and\ \citenamefont {Rickhaus}}]{deVries2024_1}%
  \BibitemOpen
  \bibfield  {author} {\bibinfo {author} {\bibfnamefont {F.~K.}\ \bibnamefont {de~Vries}}, \bibinfo {author} {\bibfnamefont {S.}~\bibnamefont {Slizovskiy}}, \bibinfo {author} {\bibfnamefont {P.}~\bibnamefont {Tomić}}, \bibinfo {author} {\bibfnamefont {R.}~\bibnamefont {Krishna~Kumar}}, \bibinfo {author} {\bibfnamefont {A.}~\bibnamefont {Garcia-Ruiz}}, \bibinfo {author} {\bibfnamefont {G.}~\bibnamefont {Zheng}}, \bibinfo {author} {\bibfnamefont {E.}~\bibnamefont {Portol{\'e}s}}, \bibinfo {author} {\bibfnamefont {L.~A.}\ \bibnamefont {Ponomarenko}}, \bibinfo {author} {\bibfnamefont {A.~K.}\ \bibnamefont {Geim}}, \bibinfo {author} {\bibfnamefont {K.}~\bibnamefont {Watanabe}}, \bibinfo {author} {\bibfnamefont {T.}~\bibnamefont {Taniguchi}}, \bibinfo {author} {\bibfnamefont {V.}~\bibnamefont {Fal’ko}}, \bibinfo {author} {\bibfnamefont {K.}~\bibnamefont {Ensslin}}, \bibinfo {author} {\bibfnamefont {T.}~\bibnamefont {Ihn}},\ and\ \bibinfo {author} {\bibfnamefont {P.}~\bibnamefont {Rickhaus}},\ }\href
  {https://doi.org/10.1021/acs.nanolett.3c03524} {\bibfield  {journal} {\bibinfo  {journal} {Nano Lett.}\ }\textbf {\bibinfo {volume} {24}},\ \bibinfo {pages} {601} (\bibinfo {year} {2024})}\BibitemShut {NoStop}%
\bibitem [{\citenamefont {Moon}\ \emph {et~al.}(2024)\citenamefont {Moon}, \citenamefont {Kim}, \citenamefont {Koshino}, \citenamefont {Taniguchi}, \citenamefont {Watanabe},\ and\ \citenamefont {Smet}}]{Moon2024_1}%
  \BibitemOpen
  \bibfield  {author} {\bibinfo {author} {\bibfnamefont {P.}~\bibnamefont {Moon}}, \bibinfo {author} {\bibfnamefont {Y.}~\bibnamefont {Kim}}, \bibinfo {author} {\bibfnamefont {M.}~\bibnamefont {Koshino}}, \bibinfo {author} {\bibfnamefont {T.}~\bibnamefont {Taniguchi}}, \bibinfo {author} {\bibfnamefont {K.}~\bibnamefont {Watanabe}},\ and\ \bibinfo {author} {\bibfnamefont {J.~H.}\ \bibnamefont {Smet}},\ }\href {https://doi.org/10.1021/acs.nanolett.3c04444} {\bibfield  {journal} {\bibinfo  {journal} {Nano Lett.}\ }\textbf {\bibinfo {volume} {24}},\ \bibinfo {pages} {3339} (\bibinfo {year} {2024})}\BibitemShut {NoStop}%
\bibitem [{\citenamefont {Bocarsly}\ \emph {et~al.}(2024)\citenamefont {Bocarsly}, \citenamefont {Uzan}, \citenamefont {Roy}, \citenamefont {Grover}, \citenamefont {Xiao}, \citenamefont {Dong}, \citenamefont {Labendik}, \citenamefont {Uri}, \citenamefont {Huber}, \citenamefont {Myasoedov}, \citenamefont {Watanabe}, \citenamefont {Taniguchi}, \citenamefont {Yan}, \citenamefont {Levitov},\ and\ \citenamefont {Zeldov}}]{bocarsly2024_1}%
  \BibitemOpen
  \bibfield  {author} {\bibinfo {author} {\bibfnamefont {M.}~\bibnamefont {Bocarsly}}, \bibinfo {author} {\bibfnamefont {M.}~\bibnamefont {Uzan}}, \bibinfo {author} {\bibfnamefont {I.}~\bibnamefont {Roy}}, \bibinfo {author} {\bibfnamefont {S.}~\bibnamefont {Grover}}, \bibinfo {author} {\bibfnamefont {J.}~\bibnamefont {Xiao}}, \bibinfo {author} {\bibfnamefont {Z.}~\bibnamefont {Dong}}, \bibinfo {author} {\bibfnamefont {M.}~\bibnamefont {Labendik}}, \bibinfo {author} {\bibfnamefont {A.}~\bibnamefont {Uri}}, \bibinfo {author} {\bibfnamefont {M.~E.}\ \bibnamefont {Huber}}, \bibinfo {author} {\bibfnamefont {Y.}~\bibnamefont {Myasoedov}}, \bibinfo {author} {\bibfnamefont {K.}~\bibnamefont {Watanabe}}, \bibinfo {author} {\bibfnamefont {T.}~\bibnamefont {Taniguchi}}, \bibinfo {author} {\bibfnamefont {B.}~\bibnamefont {Yan}}, \bibinfo {author} {\bibfnamefont {L.~S.}\ \bibnamefont {Levitov}},\ and\ \bibinfo {author} {\bibfnamefont {E.}~\bibnamefont {Zeldov}},\ }\href {https://doi.org/10.1126/science.adh3499} {\bibfield
   {journal} {\bibinfo  {journal} {Science}\ }\textbf {\bibinfo {volume} {383}},\ \bibinfo {pages} {42} (\bibinfo {year} {2024})}\BibitemShut {NoStop}%
\bibitem [{\citenamefont {Han}\ \emph {et~al.}(2025)\citenamefont {Han}, \citenamefont {Xia}, \citenamefont {Watanabe}, \citenamefont {Taniguchi}, \citenamefont {Mak},\ and\ \citenamefont {Shan}}]{han2025}%
  \BibitemOpen
  \bibfield  {author} {\bibinfo {author} {\bibfnamefont {Z.}~\bibnamefont {Han}}, \bibinfo {author} {\bibfnamefont {Y.}~\bibnamefont {Xia}}, \bibinfo {author} {\bibfnamefont {K.}~\bibnamefont {Watanabe}}, \bibinfo {author} {\bibfnamefont {T.}~\bibnamefont {Taniguchi}}, \bibinfo {author} {\bibfnamefont {K.~F.}\ \bibnamefont {Mak}},\ and\ \bibinfo {author} {\bibfnamefont {J.}~\bibnamefont {Shan}},\ }\href {https://doi.org/10.48550/arXiv.2509.19287} (\bibinfo {year} {2025}),\ \bibinfo {note} {arXiv:2509.19287 [cond-mat]}\BibitemShut {NoStop}%
\bibitem [{\citenamefont {Pippard}(1962)}]{pippard1962}%
  \BibitemOpen
  \bibfield  {author} {\bibinfo {author} {\bibfnamefont {A.~B.}\ \bibnamefont {Pippard}},\ }\href {https://doi.org/10.1098/rspa.1962.0200} {\bibfield  {journal} {\bibinfo  {journal} {Proc. R. Soc. A}\ }\textbf {\bibinfo {volume} {270}},\ \bibinfo {pages} {1} (\bibinfo {year} {1962})}\BibitemShut {NoStop}%
\bibitem [{\citenamefont {Priestley}\ \emph {et~al.}(1963)\citenamefont {Priestley}, \citenamefont {Falicov},\ and\ \citenamefont {Weisz}}]{Priestley1963}%
  \BibitemOpen
  \bibfield  {author} {\bibinfo {author} {\bibfnamefont {M.~G.}\ \bibnamefont {Priestley}}, \bibinfo {author} {\bibfnamefont {L.~M.}\ \bibnamefont {Falicov}},\ and\ \bibinfo {author} {\bibfnamefont {G.}~\bibnamefont {Weisz}},\ }\href {https://doi.org/10.1103/PhysRev.131.617} {\bibfield  {journal} {\bibinfo  {journal} {Phys. Rev.}\ }\textbf {\bibinfo {volume} {131}},\ \bibinfo {pages} {617} (\bibinfo {year} {1963})}\BibitemShut {NoStop}%
\bibitem [{\citenamefont {Watts}(1964)}]{Watts1964}%
  \BibitemOpen
  \bibfield  {author} {\bibinfo {author} {\bibfnamefont {B.~R.}\ \bibnamefont {Watts}},\ }\href {https://doi.org/10.1098/rspa.1964.0249} {\bibfield  {journal} {\bibinfo  {journal} {Proc. R. Soc. A}\ }\textbf {\bibinfo {volume} {282}},\ \bibinfo {pages} {521} (\bibinfo {year} {1964})}\BibitemShut {NoStop}%
\bibitem [{Sup()}]{SuppMat}%
  \BibitemOpen
  \href@noop {} {}\bibinfo {note} {See supplementary materials at [url] for details on device fabrication, transport measurements, band structure calculations, and additional data.}\BibitemShut {Stop}%
\bibitem [{\citenamefont {Moon}\ and\ \citenamefont {Koshino}(2014)}]{moon2014}%
  \BibitemOpen
  \bibfield  {author} {\bibinfo {author} {\bibfnamefont {P.}~\bibnamefont {Moon}}\ and\ \bibinfo {author} {\bibfnamefont {M.}~\bibnamefont {Koshino}},\ }\href {https://doi.org/10.1103/PhysRevB.90.155406} {\bibfield  {journal} {\bibinfo  {journal} {Phys. Rev. B}\ }\textbf {\bibinfo {volume} {90}},\ \bibinfo {pages} {155406} (\bibinfo {year} {2014})}\BibitemShut {NoStop}%
\bibitem [{\citenamefont {Brown}(1964)}]{Brown1964}%
  \BibitemOpen
  \bibfield  {author} {\bibinfo {author} {\bibfnamefont {E.}~\bibnamefont {Brown}},\ }\href {https://doi.org/10.1103/PhysRev.133.A1038} {\bibfield  {journal} {\bibinfo  {journal} {Phys. Rev.}\ }\textbf {\bibinfo {volume} {133}},\ \bibinfo {pages} {A1038} (\bibinfo {year} {1964})}\BibitemShut {NoStop}%
\bibitem [{\citenamefont {Zak}(1964)}]{Zak1964}%
  \BibitemOpen
  \bibfield  {author} {\bibinfo {author} {\bibfnamefont {J.}~\bibnamefont {Zak}},\ }\href {https://doi.org/10.1103/PhysRev.134.A1602} {\bibfield  {journal} {\bibinfo  {journal} {Phys. Rev.}\ }\textbf {\bibinfo {volume} {134}},\ \bibinfo {pages} {A1602} (\bibinfo {year} {1964})}\BibitemShut {NoStop}%
\bibitem [{\citenamefont {Shilov}\ \emph {et~al.}(2024)\citenamefont {Shilov}, \citenamefont {Kashchenko}, \citenamefont {Pantaleón~Peralta}, \citenamefont {Wang}, \citenamefont {Kravtsov}, \citenamefont {Kudriashov}, \citenamefont {Zhan}, \citenamefont {Taniguchi}, \citenamefont {Watanabe}, \citenamefont {Slizovskiy}, \citenamefont {Novoselov}, \citenamefont {Fal’ko}, \citenamefont {Guinea},\ and\ \citenamefont {Bandurin}}]{shilov2024}%
  \BibitemOpen
  \bibfield  {author} {\bibinfo {author} {\bibfnamefont {A.~L.}\ \bibnamefont {Shilov}}, \bibinfo {author} {\bibfnamefont {M.~A.}\ \bibnamefont {Kashchenko}}, \bibinfo {author} {\bibfnamefont {P.~A.}\ \bibnamefont {Pantaleón~Peralta}}, \bibinfo {author} {\bibfnamefont {Y.}~\bibnamefont {Wang}}, \bibinfo {author} {\bibfnamefont {M.}~\bibnamefont {Kravtsov}}, \bibinfo {author} {\bibfnamefont {A.}~\bibnamefont {Kudriashov}}, \bibinfo {author} {\bibfnamefont {Z.}~\bibnamefont {Zhan}}, \bibinfo {author} {\bibfnamefont {T.}~\bibnamefont {Taniguchi}}, \bibinfo {author} {\bibfnamefont {K.}~\bibnamefont {Watanabe}}, \bibinfo {author} {\bibfnamefont {S.}~\bibnamefont {Slizovskiy}}, \bibinfo {author} {\bibfnamefont {K.~S.}\ \bibnamefont {Novoselov}}, \bibinfo {author} {\bibfnamefont {V.~I.}\ \bibnamefont {Fal’ko}}, \bibinfo {author} {\bibfnamefont {F.}~\bibnamefont {Guinea}},\ and\ \bibinfo {author} {\bibfnamefont {D.~A.}\ \bibnamefont {Bandurin}},\ }\href {https://doi.org/10.1021/acsnano.3c13212} {\bibfield
  {journal} {\bibinfo  {journal} {ACS Nano}\ }\textbf {\bibinfo {volume} {18}},\ \bibinfo {pages} {11769} (\bibinfo {year} {2024})}\BibitemShut {NoStop}%
\bibitem [{\citenamefont {Wang}\ \emph {et~al.}()\citenamefont {Wang}, \citenamefont {Perez-Piskunow}, \citenamefont {Wong}, \citenamefont {Holwill}, \citenamefont {Liu}, \citenamefont {Fu}, \citenamefont {Hu}, \citenamefont {Taniguchi}, \citenamefont {Watanabe}, \citenamefont {Ariando}, \citenamefont {Li}, \citenamefont {Goh}, \citenamefont {Roche}, \citenamefont {Jung}, \citenamefont {Novoselov},\ and\ \citenamefont {Leconte}}]{wang2023openorbitinducedlowfield}%
  \BibitemOpen
  \bibfield  {author} {\bibinfo {author} {\bibfnamefont {Z.}~\bibnamefont {Wang}}, \bibinfo {author} {\bibfnamefont {P.~M.}\ \bibnamefont {Perez-Piskunow}}, \bibinfo {author} {\bibfnamefont {C.~P.~Y.}\ \bibnamefont {Wong}}, \bibinfo {author} {\bibfnamefont {M.}~\bibnamefont {Holwill}}, \bibinfo {author} {\bibfnamefont {J.}~\bibnamefont {Liu}}, \bibinfo {author} {\bibfnamefont {W.}~\bibnamefont {Fu}}, \bibinfo {author} {\bibfnamefont {J.}~\bibnamefont {Hu}}, \bibinfo {author} {\bibfnamefont {T.}~\bibnamefont {Taniguchi}}, \bibinfo {author} {\bibfnamefont {K.}~\bibnamefont {Watanabe}}, \bibinfo {author} {\bibfnamefont {A.}~\bibnamefont {Ariando}}, \bibinfo {author} {\bibfnamefont {L.}~\bibnamefont {Li}}, \bibinfo {author} {\bibfnamefont {K.~E.~J.}\ \bibnamefont {Goh}}, \bibinfo {author} {\bibfnamefont {S.}~\bibnamefont {Roche}}, \bibinfo {author} {\bibfnamefont {J.}~\bibnamefont {Jung}}, \bibinfo {author} {\bibfnamefont {K.}~\bibnamefont {Novoselov}},\ and\ \bibinfo {author} {\bibfnamefont {N.}~\bibnamefont
  {Leconte}},\ }\href {https://arxiv.org/abs/2312.07004} \Eprint {https://arxiv.org/abs/2312.07004} {arXiv:2312.07004} \BibitemShut {NoStop}%
\bibitem [{\citenamefont {Kemble}(1935)}]{Kemble1935}%
  \BibitemOpen
  \bibfield  {author} {\bibinfo {author} {\bibfnamefont {E.~C.}\ \bibnamefont {Kemble}},\ }\href {https://doi.org/10.1103/PhysRev.48.549} {\bibfield  {journal} {\bibinfo  {journal} {Phys. Rev.}\ }\textbf {\bibinfo {volume} {48}},\ \bibinfo {pages} {549} (\bibinfo {year} {1935})}\BibitemShut {NoStop}%
\bibitem [{\citenamefont {Patri}\ and\ \citenamefont {Franz}(2025)}]{Patri2025}%
  \BibitemOpen
  \bibfield  {author} {\bibinfo {author} {\bibfnamefont {A.~S.}\ \bibnamefont {Patri}}\ and\ \bibinfo {author} {\bibfnamefont {M.}~\bibnamefont {Franz}},\ }\href {https://doi.org/10.1103/pgqh-wm5v} {\bibfield  {journal} {\bibinfo  {journal} {Phys. Rev. B}\ }\textbf {\bibinfo {volume} {112}},\ \bibinfo {pages} {214505} (\bibinfo {year} {2025})}\BibitemShut {NoStop}%
\bibitem [{\citenamefont {Sheekey}\ \emph {et~al.}()\citenamefont {Sheekey}, \citenamefont {Arp}, \citenamefont {Foutty}, \citenamefont {Zhang}, \citenamefont {Tan}, \citenamefont {Holleis}, \citenamefont {Guo}, \citenamefont {Kalantre}, \citenamefont {Zhang}, \citenamefont {Zakharyan}, \citenamefont {Gong}, \citenamefont {Keough}, \citenamefont {Choi}, \citenamefont {Choi}, \citenamefont {Xu}, \citenamefont {Xie}, \citenamefont {Alexander}, \citenamefont {Hocking}, \citenamefont {Cao}, \citenamefont {Huber}, \citenamefont {Taniguchi}, \citenamefont {Watanabe}, \citenamefont {Jin}, \citenamefont {Lantagne-Hurtubise}, \citenamefont {Sharpe}, \citenamefont {Devakul},\ and\ \citenamefont {Young}}]{Sheekey2026}%
  \BibitemOpen
  \bibfield  {author} {\bibinfo {author} {\bibfnamefont {O.~I.}\ \bibnamefont {Sheekey}}, \bibinfo {author} {\bibfnamefont {T.~B.}\ \bibnamefont {Arp}}, \bibinfo {author} {\bibfnamefont {B.~A.}\ \bibnamefont {Foutty}}, \bibinfo {author} {\bibfnamefont {R.}~\bibnamefont {Zhang}}, \bibinfo {author} {\bibfnamefont {T.}~\bibnamefont {Tan}}, \bibinfo {author} {\bibfnamefont {L.~F.~W.}\ \bibnamefont {Holleis}}, \bibinfo {author} {\bibfnamefont {Y.}~\bibnamefont {Guo}}, \bibinfo {author} {\bibfnamefont {S.~S.}\ \bibnamefont {Kalantre}}, \bibinfo {author} {\bibfnamefont {C.}~\bibnamefont {Zhang}}, \bibinfo {author} {\bibfnamefont {M.}~\bibnamefont {Zakharyan}}, \bibinfo {author} {\bibfnamefont {D.}~\bibnamefont {Gong}}, \bibinfo {author} {\bibfnamefont {A.}~\bibnamefont {Keough}}, \bibinfo {author} {\bibfnamefont {Y.}~\bibnamefont {Choi}}, \bibinfo {author} {\bibfnamefont {Y.}~\bibnamefont {Choi}}, \bibinfo {author} {\bibfnamefont {S.}~\bibnamefont {Xu}}, \bibinfo {author} {\bibfnamefont {T.}~\bibnamefont {Xie}},
  \bibinfo {author} {\bibfnamefont {B.~H.}\ \bibnamefont {Alexander}}, \bibinfo {author} {\bibfnamefont {M.}~\bibnamefont {Hocking}}, \bibinfo {author} {\bibfnamefont {Q.}~\bibnamefont {Cao}}, \bibinfo {author} {\bibfnamefont {M.~E.}\ \bibnamefont {Huber}}, \bibinfo {author} {\bibfnamefont {T.}~\bibnamefont {Taniguchi}}, \bibinfo {author} {\bibfnamefont {K.}~\bibnamefont {Watanabe}}, \bibinfo {author} {\bibfnamefont {C.}~\bibnamefont {Jin}}, \bibinfo {author} {\bibfnamefont {E.}~\bibnamefont {Lantagne-Hurtubise}}, \bibinfo {author} {\bibfnamefont {A.}~\bibnamefont {Sharpe}}, \bibinfo {author} {\bibfnamefont {T.}~\bibnamefont {Devakul}},\ and\ \bibinfo {author} {\bibfnamefont {A.~F.}\ \bibnamefont {Young}},\ }\href {https://arxiv.org/abs/2605.30316}  \Eprint {https://arxiv.org/abs/2605.30316} {arXiv:2605.30316} \BibitemShut {NoStop}%
\bibitem [{\citenamefont {Krishna~Kumar}\ \emph {et~al.}(2017)\citenamefont {Krishna~Kumar}, \citenamefont {Chen}, \citenamefont {Auton}, \citenamefont {Mishchenko}, \citenamefont {Bandurin}, \citenamefont {Morozov}, \citenamefont {Cao}, \citenamefont {Khestanova}, \citenamefont {Ben~Shalom}, \citenamefont {Kretinin}, \citenamefont {Novoselov}, \citenamefont {Eaves}, \citenamefont {Grigorieva}, \citenamefont {Ponomarenko}, \citenamefont {Fal’ko},\ and\ \citenamefont {Geim}}]{krishnakumar2017}%
  \BibitemOpen
  \bibfield  {author} {\bibinfo {author} {\bibfnamefont {R.}~\bibnamefont {Krishna~Kumar}}, \bibinfo {author} {\bibfnamefont {X.}~\bibnamefont {Chen}}, \bibinfo {author} {\bibfnamefont {G.~H.}\ \bibnamefont {Auton}}, \bibinfo {author} {\bibfnamefont {A.}~\bibnamefont {Mishchenko}}, \bibinfo {author} {\bibfnamefont {D.~A.}\ \bibnamefont {Bandurin}}, \bibinfo {author} {\bibfnamefont {S.~V.}\ \bibnamefont {Morozov}}, \bibinfo {author} {\bibfnamefont {Y.}~\bibnamefont {Cao}}, \bibinfo {author} {\bibfnamefont {E.}~\bibnamefont {Khestanova}}, \bibinfo {author} {\bibfnamefont {M.}~\bibnamefont {Ben~Shalom}}, \bibinfo {author} {\bibfnamefont {A.~V.}\ \bibnamefont {Kretinin}}, \bibinfo {author} {\bibfnamefont {K.~S.}\ \bibnamefont {Novoselov}}, \bibinfo {author} {\bibfnamefont {L.}~\bibnamefont {Eaves}}, \bibinfo {author} {\bibfnamefont {I.~V.}\ \bibnamefont {Grigorieva}}, \bibinfo {author} {\bibfnamefont {L.~A.}\ \bibnamefont {Ponomarenko}}, \bibinfo {author} {\bibfnamefont {V.~I.}\ \bibnamefont {Fal’ko}},\ and\
  \bibinfo {author} {\bibfnamefont {A.~K.}\ \bibnamefont {Geim}},\ }\href {https://doi.org/10.1126/science.aal3357} {\bibfield  {journal} {\bibinfo  {journal} {Science}\ }\textbf {\bibinfo {volume} {357}},\ \bibinfo {pages} {181} (\bibinfo {year} {2017})}\BibitemShut {NoStop}%
\bibitem [{\citenamefont {{Huber}}\ \emph {et~al.}(2023)\citenamefont {{Huber}}, \citenamefont {{Leeb}}, \citenamefont {{Bauer}}, \citenamefont {{Benka}}, \citenamefont {{Knolle}}, \citenamefont {{Pfleiderer}},\ and\ \citenamefont {{Wilde}}}]{Huber2023}%
  \BibitemOpen
  \bibfield  {author} {\bibinfo {author} {\bibfnamefont {N.}~\bibnamefont {{Huber}}}, \bibinfo {author} {\bibfnamefont {V.}~\bibnamefont {{Leeb}}}, \bibinfo {author} {\bibfnamefont {A.}~\bibnamefont {{Bauer}}}, \bibinfo {author} {\bibfnamefont {G.}~\bibnamefont {{Benka}}}, \bibinfo {author} {\bibfnamefont {J.}~\bibnamefont {{Knolle}}}, \bibinfo {author} {\bibfnamefont {C.}~\bibnamefont {{Pfleiderer}}},\ and\ \bibinfo {author} {\bibfnamefont {M.~A.}\ \bibnamefont {{Wilde}}},\ }\href {https://doi.org/10.1038/s41586-023-06330-y} {\bibfield  {journal} {\bibinfo  {journal} {\nat}\ }\textbf {\bibinfo {volume} {621}},\ \bibinfo {pages} {276} (\bibinfo {year} {2023})}\BibitemShut {NoStop}%
\bibitem [{\citenamefont {Leeb}\ \emph {et~al.}(2025)\citenamefont {Leeb}, \citenamefont {Huber}, \citenamefont {Pfleiderer}, \citenamefont {Knolle},\ and\ \citenamefont {Wilde}}]{Leeb2025}%
  \BibitemOpen
  \bibfield  {author} {\bibinfo {author} {\bibfnamefont {V.}~\bibnamefont {Leeb}}, \bibinfo {author} {\bibfnamefont {N.}~\bibnamefont {Huber}}, \bibinfo {author} {\bibfnamefont {C.}~\bibnamefont {Pfleiderer}}, \bibinfo {author} {\bibfnamefont {J.}~\bibnamefont {Knolle}},\ and\ \bibinfo {author} {\bibfnamefont {M.~A.}\ \bibnamefont {Wilde}},\ }\href {https://doi.org/https://doi.org/10.1002/apxr.202400134} {\bibfield  {journal} {\bibinfo  {journal} {Adv. Phys. Res.}\ }\textbf {\bibinfo {volume} {4}},\ \bibinfo {pages} {2400134} (\bibinfo {year} {2025})}\BibitemShut {NoStop}%
\bibitem [{\citenamefont {Leadley}\ \emph {et~al.}(1992)\citenamefont {Leadley}, \citenamefont {Fletcher}, \citenamefont {Nicholas}, \citenamefont {Tao}, \citenamefont {Foxon},\ and\ \citenamefont {Harris}}]{leadley1992}%
  \BibitemOpen
  \bibfield  {author} {\bibinfo {author} {\bibfnamefont {D.~R.}\ \bibnamefont {Leadley}}, \bibinfo {author} {\bibfnamefont {R.}~\bibnamefont {Fletcher}}, \bibinfo {author} {\bibfnamefont {R.~J.}\ \bibnamefont {Nicholas}}, \bibinfo {author} {\bibfnamefont {F.}~\bibnamefont {Tao}}, \bibinfo {author} {\bibfnamefont {C.~T.}\ \bibnamefont {Foxon}},\ and\ \bibinfo {author} {\bibfnamefont {J.~J.}\ \bibnamefont {Harris}},\ }\href {https://doi.org/10.1103/physrevb.46.12439} {\bibfield  {journal} {\bibinfo  {journal} {Physical Review. B, Condensed Matter}\ }\textbf {\bibinfo {volume} {46}},\ \bibinfo {pages} {12439} (\bibinfo {year} {1992})}\BibitemShut {NoStop}%
\bibitem [{\citenamefont {Raikh}\ and\ \citenamefont {Shahbazyan}(1994)}]{raikh1994}%
  \BibitemOpen
  \bibfield  {author} {\bibinfo {author} {\bibfnamefont {M.~E.}\ \bibnamefont {Raikh}}\ and\ \bibinfo {author} {\bibfnamefont {T.~V.}\ \bibnamefont {Shahbazyan}},\ }\href {https://doi.org/10.1103/PhysRevB.49.5531} {\bibfield  {journal} {\bibinfo  {journal} {Physical Review B}\ }\textbf {\bibinfo {volume} {49}},\ \bibinfo {pages} {5531} (\bibinfo {year} {1994})}\BibitemShut {NoStop}%
\bibitem [{\citenamefont {Phinney}\ \emph {et~al.}(2021)\citenamefont {Phinney}, \citenamefont {Bandurin}, \citenamefont {Collignon}, \citenamefont {Dmitriev}, \citenamefont {Taniguchi}, \citenamefont {Watanabe},\ and\ \citenamefont {Jarillo-Herrero}}]{Phinney2021}%
  \BibitemOpen
  \bibfield  {author} {\bibinfo {author} {\bibfnamefont {I.~Y.}\ \bibnamefont {Phinney}}, \bibinfo {author} {\bibfnamefont {D.~A.}\ \bibnamefont {Bandurin}}, \bibinfo {author} {\bibfnamefont {C.}~\bibnamefont {Collignon}}, \bibinfo {author} {\bibfnamefont {I.~A.}\ \bibnamefont {Dmitriev}}, \bibinfo {author} {\bibfnamefont {T.}~\bibnamefont {Taniguchi}}, \bibinfo {author} {\bibfnamefont {K.}~\bibnamefont {Watanabe}},\ and\ \bibinfo {author} {\bibfnamefont {P.}~\bibnamefont {Jarillo-Herrero}},\ }\href {https://doi.org/10.1103/PhysRevLett.127.056802} {\bibfield  {journal} {\bibinfo  {journal} {Phys. Rev. Lett.}\ }\textbf {\bibinfo {volume} {127}},\ \bibinfo {pages} {056802} (\bibinfo {year} {2021})}\BibitemShut {NoStop}%
\bibitem [{Note1()}]{Note1}%
  \BibitemOpen
  \bibinfo {note} {Increased temperature smears the FS, and QOs with cyclotron mass $m^* = \hbar ^{-2} \dv *{A_k}{\mu }$ (with $A_k$ the extremal FS cross-section area) are damped according to the well-known Lifshitz-Kosevich thermal damping factor $R_T = X/\sinh {X}$, with $X\propto m^* T / B$ \cite {Lifshitz}. Density-independent QO frequencies, with near-zero $m^*$, can survive to much higher temperatures~\cite {deVries2024_1,Leeb2025}.}\BibitemShut {Stop}%
\bibitem [{\citenamefont {Lu}\ and\ \citenamefont {Fertig}(2014)}]{lu2014}%
  \BibitemOpen
  \bibfield  {author} {\bibinfo {author} {\bibfnamefont {C.-K.}\ \bibnamefont {Lu}}\ and\ \bibinfo {author} {\bibfnamefont {H.~A.}\ \bibnamefont {Fertig}},\ }\href {https://doi.org/10.1103/PhysRevB.89.085408} {\bibfield  {journal} {\bibinfo  {journal} {Physical Review B}\ }\textbf {\bibinfo {volume} {89}},\ \bibinfo {pages} {085408} (\bibinfo {year} {2014})}\BibitemShut {NoStop}%
\bibitem [{\citenamefont {Kalantre}\ \emph {et~al.}(2026)\citenamefont {Kalantre}, \citenamefont {Alexander}, \citenamefont {May-Mann}, \citenamefont {Herzog-Arbeitman}, \citenamefont {Hocking}, \citenamefont {Cao}, \citenamefont {Watanabe}, \citenamefont {Taniguchi}, \citenamefont {Goldhaber-Gordon}, \citenamefont {Mannix}, \citenamefont {Devakul}, \citenamefont {Kwan}, \citenamefont {Parker},\ and\ \citenamefont {Sharpe}}]{kalantre2026}%
  \BibitemOpen
  \bibfield  {author} {\bibinfo {author} {\bibfnamefont {S.~S.}\ \bibnamefont {Kalantre}}, \bibinfo {author} {\bibfnamefont {B.~H.}\ \bibnamefont {Alexander}}, \bibinfo {author} {\bibfnamefont {J.}~\bibnamefont {May-Mann}}, \bibinfo {author} {\bibfnamefont {J.}~\bibnamefont {Herzog-Arbeitman}}, \bibinfo {author} {\bibfnamefont {M.}~\bibnamefont {Hocking}}, \bibinfo {author} {\bibfnamefont {Q.}~\bibnamefont {Cao}}, \bibinfo {author} {\bibfnamefont {K.}~\bibnamefont {Watanabe}}, \bibinfo {author} {\bibfnamefont {T.}~\bibnamefont {Taniguchi}}, \bibinfo {author} {\bibfnamefont {D.}~\bibnamefont {Goldhaber-Gordon}}, \bibinfo {author} {\bibfnamefont {A.~J.}\ \bibnamefont {Mannix}}, \bibinfo {author} {\bibfnamefont {T.}~\bibnamefont {Devakul}}, \bibinfo {author} {\bibfnamefont {Y.~H.}\ \bibnamefont {Kwan}}, \bibinfo {author} {\bibfnamefont {D.~E.}\ \bibnamefont {Parker}},\ and\ \bibinfo {author} {\bibfnamefont {A.}~\bibnamefont {Sharpe}},\ }\href {https://doi.org/10.48550/arXiv.2606.05356} \bibinfo {note} {arXiv:2606.05356 [cond-mat.supr-con]}\BibitemShut {NoStop}%
\bibitem [{\citenamefont {Zhao}\ \emph {et~al.}(2026)\citenamefont {Zhao}, \citenamefont {Chou},\ and\ \citenamefont {Sarma}}]{zhao2026}%
  \BibitemOpen
  \bibfield  {author} {\bibinfo {author} {\bibfnamefont {J.-Y.}\ \bibnamefont {Zhao}}, \bibinfo {author} {\bibfnamefont {Y.-Z.}\ \bibnamefont {Chou}},\ and\ \bibinfo {author} {\bibfnamefont {S.~D.}\ \bibnamefont {Sarma}},\ }\href {https://doi.org/10.48550/arXiv.2607.27207}  \bibinfo {note} {arXiv:2607.27207 [cond-mat.mes-hall]}\BibitemShut {NoStop}%
\bibitem [{\citenamefont {{Lifshitz}}\ and\ \citenamefont {{Kosevich}}(1956)}]{Lifshitz}%
  \BibitemOpen
  \bibfield  {author} {\bibinfo {author} {\bibfnamefont {I.~M.}\ \bibnamefont {{Lifshitz}}}\ and\ \bibinfo {author} {\bibfnamefont {A.~M.}\ \bibnamefont {{Kosevich}}},\ }\href {https://jetp.ras.ru/cgi-bin/e/index/e/2/4/p636?a=list} {\bibfield  {journal} {\bibinfo  {journal} {Sov. Phys. JETP}\ }\textbf {\bibinfo {volume} {2}} (\bibinfo {year} {1956})}\BibitemShut {NoStop}%
\end{thebibliography}
\end{document}